\documentclass{mem}
\usepackage{natbib}\usepackage{txfonts}\usepackage{balance}
\usepackage{graphicx}
\usepackage[a4paper]{hyperref}
\idline{75}{282}
\begin{document}
\def\teff{$T\rm_{eff }$}
\def\kms{$\mathrm {km s}^{-1}$}

\title{
Gaia: unravelling the chemical and dinamical history of our Galaxy
}

   \subtitle{}

\author{
E. \,Pancino\inst{1} 
          }

  \offprints{P. Bonifacio}

\institute{
Istituto Nazionale di Astrofisica --
Osservatorio Astronomico di Bologna, Via Ranzani 1,
I-40127 Bologna, Italy.
\email{elena.pancino@oabo.inaf.it}
}

\authorrunning{Pancino}

\titlerunning{Gaia}

\abstract{

The Gaia astrometric mission -- the Hipparcos successor -- is described in some
detail, with its three instruments: the two (spectro)photometers (BP and RP)
covering the range 330-1050 nm, the white light (G-band) imager dedicated to
astrometry, and the radial velocity spectrometer (RVS) covering the range
847-874 nm at a resolution R$\simeq$11500. The whole sky will be scanned
repeatedly providing data for $\sim$10$^9$ point-like objects, down to a
magnitude of V$\simeq$20, aiming to the full 6D reconstruction of the Milky Way
kinematical and dinamical structure with unprecendented precision. The horizon
of scientific questions that can find an answer with such a set of data is vast,
including besides the Galaxy: Solar system studies, stellar astrophysics,
exoplanets, supernovae, Local group physics, unresolved galaxies, Quasars, and
fundamental physics. The Italian involvement in the mission preparation is
briefly outlined.

}
\maketitle{}

\section{The Gaia mission}

Gaia is a cornerstone mission of the ESA Space Program, presently scheduled for
launch in 2012. The Gaia satellite will perform an all-sky survey to obtain
parallaxes and proper motions to $\mu as$ precision for about 10$^9$ point-like
sources and astrophysical parameters ($T_{\mathrm{eff}}$, $\log g$, $E(B-V)$,
metallicity etc.) for stars down to a limiting magnitude of $V\simeq 20$, plus
2-30 km/s  accuracy (depending on spectral type), radial velocities for several
millions of stars down to $V < 17$. 

Such an observational effort has been compared to the mapping of the human genome
for the amount of collected data and for the impact that it will have on all
branches of astronomy and astrophysics. The expected end-of-mission astrometric
accuracies are almost 100 times better than the HIPPARCOS dataset
\citep[see][]{perryman97}. This exquisite precision will allow a full and
detailed reconstruction of the 3D spatial structure and 3D velocity field of the
Milky Way galaxy within $\simeq 10$ kpc from the Sun. This will provide answers
to long-standing questions about the origin and evolution of our Galaxy, from a
quantitative census of its stellar populations, to a detailed characterization of
its substructures (as, for instance, tidal streams in the Halo,
\citep[see][]{ibata07}, to the distribution of dark matter. 

The accurate 3D motion of more distant Galactic satellites (as distant globular
clusters and the Magellanic Clouds) will be also obtained by averaging the proper
motions of many thousands of member stars: this will provide an unprecedented
leverage to constrain the mass distribution of the Galaxy and/or non-standard
theories of gravitation. Gaia will determine direct geometric distances to
essentially any kind of standard candle currently used for distance
determination, setting the whole cosmological distance scale on extremely firm
bases. 

As challenging as it is, the processing and analysis of the huge data-flow
incoming from Gaia is the subject of thorough study and preparatory work by the
Data Processing and Analysis Consortium (DPAC), in charge of all aspects of the
Gaia data reduction. The consortium comprises more than 400 scientists from 25
European institutes. 

In the next Sections, I will describe in some detail the instrument and its
capabilities, including a short review of its expected scientific output, its
concept and instruments, the organization of data analysis and the Italian
contribution to the mission.

\subsection{Science goals and capabilities}

\begin{table}[t]
\caption{Expected numbers of specific objects observed by Gaia.}
\label{tab_numbers}
\begin{center}
\begin{tabular}{l r}
\hline
Type & Numbers\\
\hline
Extragalactic supernovae & 20\,000 \\ 
Resolved galaxies & 10$^6$--10$^7$ \\ 
Quasars & 500\,000 \\                  
Solar system objects & 250\,000 \\    
Brown dwarfs & $\geq$50\,000    \\    
Extra-solar planets & 15\,000\\ 
Disk white dwarfs & 200\,000\\
Astrometric microlensing events & 100 \\ 
Photometric microlensing events & 1000 \\ 
Resolved binaries (within 250 pc) & 10$^7$ \\ 
\hline
\end{tabular}
\end{center}
\end{table}

For many years, the state of the art in celestial cartography has been the
Schmidt surveys of Palomar and ESO, and their digitized counterparts. As will
bee detailed in the following Sections, the expected precision of Gaia
measurements is unprecedented, and the resulting scientific harvest will be of
almost inconceivable extent and implication.  

Gaia will provide detailed  information on stellar evolution and star formation
in our Galaxy. It will clarify the origin and formation history  of our Galaxy.
The data will enable to precisely identify relics of tidally-disrupted accretion
debris, probe the distribution of dark matter, establish the luminosity function
for pre-main sequence stars, detect and categorize rapid evolutionary stellar
phases, place unprecedented constraints on the age, internal structure and
evolution of all stellar types, establish a rigorous distance scale framework
throughout the Galaxy and beyond, and classify star formation and kinematical
and dynamical behaviour within the Local Group of galaxies. 

Gaia will pinpoint exotic objects in colossal and almost unimaginable numbers:
many thousands of extra-solar  planets will be discovered (from both their
astrometric wobble and photometric transits) and their detailed orbits and masses
determined; tens of thousands of brown dwarfs and white dwarfs will be
identified; tens of  thousands of extragalactic supernovae will be discovered;
Solar System studies will receive a massive impetus through the observation of
hundreds of thousands of minor planets; near-Earth objects, inner Trojans and
even  new trans-Neptunian objects, including Plutinos, may be discovered. 

Gaia will follow the bending of star light by the Sun and major planets over the
entire celestial sphere, and  therefore directly observe the structure of
space-time -- the accuracy of its measurement of General Relativistic light
bending may reveal the long-sought scalar correction to its tensor form. The PPN
parameters $\gamma$ and $\beta$, and the solar quadrupole moment J2, will be
determined with unprecedented precision. All this, and more, through the accurate
measurement of star positions. 

We summarize some of the most interesting object classes that will be observed by
Gaia, with estimates of the expected total number of objects, in
Table~\ref{tab_numbers}. For more information on the Gaia mission:
http://www.rssd.esa.int/Gaia. More information for the public on Gaia and its
science capabilities are contained in the {\em Gaia information
sheets}\footnote{http://www.rssd.esa.int/index.php?project=GAI
A\&page=Info\_sheets\_overview.}. An excellent review of the science
possibilities opened by Gaia can be found in \citet{perryman97}.

\subsection{Launch, timeline and data releases}

The first idea for Gaia began circulating in the early 1990, culminating in a
proposal for a  cornerstone mission within ESA's science programme submitted in
1993, and a workshop in Cambridge in June 1995. By the time the final catalogue
will be released approximately in 2020, almost two decades of work will have
elapsed between the orginal concept and mission completion.  

Gaia will be launched by a Soyuz carrier (rather than the initially foreseen
Ariane 5) in 2012 from French Guyana and will start operating once it will reach
its Lissajus orbit around L2 (the unstable Langrange point of the Sun and
Eart-Moon system), in about one month. Two ground stations will receive the
compressed Gaia data during the 5 years\footnote{If -- after careful evaluation --
the scientific output of the mission will benefit from an extension of the
operation period, the satellite should be able to gather data for one more year,
remaining within the Earth eclipse.} of operation: Cebreros (Spain) and Perth
(Australia). The data will then be transmitted to the main data centers throughout
Europe to allow for data processing. We are presently in technical development
phase C/D, and the hardware is being built, tested and assembled. Software
development started in 2006 and is presently producing and testing pipelines with
the aim of delivering to the astrophysical community a full catalogue and dataset
ready for scientific investigation (see Section~\ref{sec-dpac}).

Apart from the end-of-mission data release, foreseen around 2020, some
intermediate data releases are foreseen. In particular, there should be one first
intermediate release covering either the first 6 months or the first year of
operation, followed by a second and possibly a third intermediate release, that
are presently being discussed. The data analysis will proceed in parallel with
observations, the major pipelines re-processing all the data every 6 months, with
secondary cycle pipelines -- dedicated to specific tasks -- operating on different
timescales. In particular, verified science alerts, based on unexpected
variability in flux and/or radial velocity, are expected to be released within 24
hours from detection, after an initial period of testing and fine-tuning of the
detection algorithms. 

\begin{figure}
\resizebox{\hsize}{!}{\includegraphics{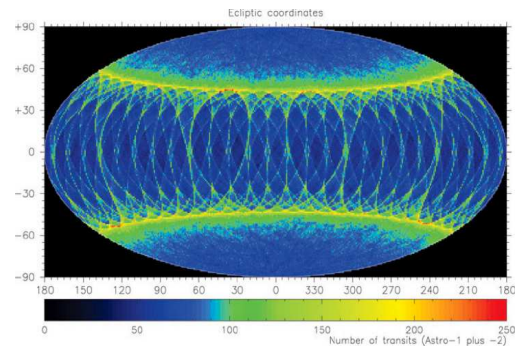}}
\caption{The average number of passages on the sky, in ecliptic coordinates.
\copyright ESA}
\label{pancino_fig_scan}
\end{figure}

\subsection{Mission concepts}

\begin{figure*}
\resizebox{\hsize}{!}{\includegraphics{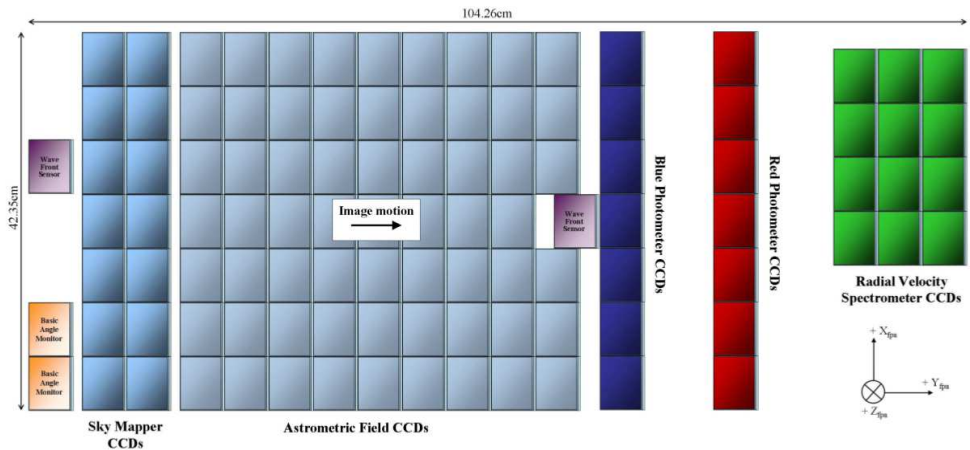}}
\caption{The Gaia focal plane. \copyright ESA}
\label{pancino_fig_foc}
\end{figure*}

During its 5-year operational lifetime, the satellite will continuously spin
around its axis, with a constant speed of 60~arcsec/sec. As a result, over a
period of 6 hours, the two astrometric fields of view will scan across all
objects located along the great circle perpendicular to the spin axis. As a
result of the basic angle of 106.5$^{\rm o}$ separating the astrometric fields of
view on the sky, objects transit the second field of view with a delay of 106.5
minutes compared to the first field. Gaia's spin axis does not point to a fixed
direction in space, but is carefully controlled so as to precess slowly on the
sky. As a result, the great circle that is mapped by the two fields of view every
6 hours changes slowly with time, allowing reapeated full sky coverage over the
mission lifetime. The best strategy, dictated by thermal stability and power
requirements, is to let the spin axis precess (with a period of 63 days) around
the solar direction with a fixed angle of 45$^{\rm o}$. The above scanning
strategy, referred to as ``revolving scanning", was successfully adopted during
the Hipparcos mission. 

Every sky region will be scanned on average 70-80 times, with regions lying at
$\pm$45$^{\rm o}$ from the Ecliptic Poles being scanned on average more often
than other locations. Each of the Gaia targets will be therefore scanned (within
differently inclined great circles) from a minimum of approximately 10 times to a
maximum of 250 times (Figure~\ref{pancino_fig_scan}). Only point-like sources
will be observed, and in some regions of the sky, like the Baade's window,
$\omega$ Centauri or other globular clusters, the star density of the two
combined fields of view will be of the order of 750\,000 or more per square
degree, exceeding the storage capability of the onboard processors, so Gaia will
not study in detail these dense areas. 

\subsection{Focal plane}

Figure~\ref{pancino_fig_foc} shows the focal plane of Gaia, with its 105 CCDs,
which are read in TDI (Time Delayed Integration) mode. Objects enter the focal
plane from the left and cross one CCD in 4 seconds. Apart from some technical
CCDs that are of little interest in this context, the first two CCD columns, the
Sky Mappers (SM), perform the on-board detection of point-like sources, each of
the two columns being able to see only one of the two lines of sight. After the
objects are identified and selected, small windows are assigned, which follow
them in the astrometric field (AF) CCDs where white light (or G-band) images are
obtained (Section~\ref{sec-af}). Immediately following the AF, two additional
columns of CCDs gather light from two slitless prism spectrographs, the blue
spectrophotometer (BP) and the red one (RP), which produce dispersed images
(Section~\ref{sec-phot}). Finally, objects transit on the Radial Velocity
Spectrometer (RVS) CCDs to produce higher resolution spectra around the Calcium
Triplet (CaT) region (Section~\ref{sec-rvs}). 

\subsection{Astrometry} 
\label{sec-af}

\begin{figure}
\resizebox{\hsize}{!}{\includegraphics{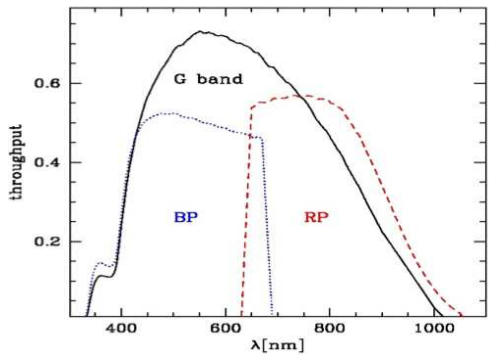}}
\caption{The passbands of the G-band, BP and RP. \copyright ESA} 
\label{pancino_fig_band}
\end{figure}

The AF CCDs will provide G-band images, i.e., white light images where the
passband is defined by the telescope optics transmission and the CCDs sensitivity,
with a very broad combined passband ranging from 330 to 1050~nm and peaking around
500--600~nm (Figure~\ref{pancino_fig_band}). The objective of Gaia's astrometric
data reduction system is the construction of core mission products: the five
standard astrometric parameters, position ($\alpha$, $\delta$), parallax
($\varpi$), and proper motion ($\mu_{\alpha}$, $\mu_{\delta}$) for all observed
stellar objects. The expected end-of-mission precision in the proper motions is
expected to be better than 10~$\mu$as for G$<$10 stars, 25~$\mu$as for G=15, and
300~$\mu$as for G=20. For parallaxes, considering a G=12 star, we can expect to
have distances at better than 0.1\% within 250~pc, 1\% within 2700~pc, and 10\%
within 10~kpc.

To reach these end-of-mission precisions, the average 70--80 observations per
target gathered during the 5-year mission duration will have to be combined into a
single, global, and self-consistent manner. 40~Gb of telemetry data will first
pass through the Initial Data Treatment (IDT) which determines the image
parameters and centroids, and then performes an object cross-matching. The output
forms the so-called One Day Astrometric Solution (ODAS), together with the
satellite attitude and calibration, to the sub-milliarcsecond accuracy.  The data
are then written to the Main Database. 

The next step is the Astrometric Global Iterative Solution (AGIS) processing. AGIS
processes together the attitude and calibration parameters with  the source
parameters, refining them in an iterative procedure that stops when the
adjustments become sufficiently small. As soon as new data come in, on the basis
of 6 months cycles, all the data in hand are reprocessed toghether from scratch.
This is the only scheme that allows for the quoted precisions, and it is also the
philosophy that justifies Gaia as a self-calibrationg mission. The primary AGIS
cycle will treat only stars that are flagged as single and non-variable (expected
to be around 500 millions), while other kinds of objects will be computed in
secondary AGIS cycles that utilize the main AGIS sulution. Dedicated pipelines for
specific kinds of objects (asteroids, slightly extended objects, variable objects
and so on) are being put in place to extract the best possible precision. Owing to
the large data volume (100~Tb) that Gaia will produce, and to the iterative nature
of the processing, the computing challenges are formidable: AGIS processing alone
requires some 10$^{21}$~FLOPs which translates to runtimes of months on the ESAC
computers in Madrid.

\subsection{Spectrophotometry}
\label{sec-phot}

The primary aim of the photometric instrument is mission critical in two respects:
(i) to correct the measured centroids position in the AF for systematic chromatic
effects, and (ii) to classify and determine astrophysical characteristics of all
objects, such as temperature, gravity, mass, age and chemical composition (in the
case of stars).

The BP and RP spectrophotometers are based on a dispersive-prism approach such
that the incoming light is not focussed in a PSF-like spot, but dispersed along
the scan direction in a low-resolution spectrum.
The BP operates between 330--680~nm while the RP between 640-1000~nm
(Figure~\ref{pancino_fig_band}). Both prisms have appropriate broad-band filters
to block unwanted light. The two dedicated CCD stripes cover the full height of
the AF and, therefore, all objects that are imaged in the AF are also imaged in
the BP and RP. 

\begin{figure}
\resizebox{\hsize}{!}{\includegraphics{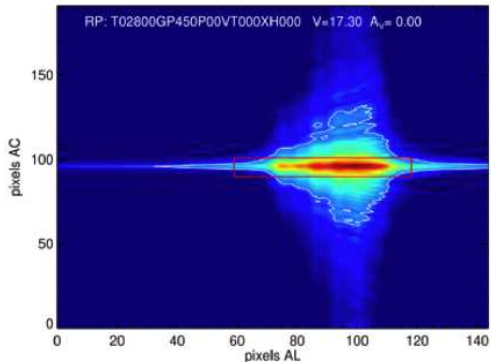}}
\caption{A simulated RP dispersed image, with a red rectangle marking the window
assigned for compression and ground telemetry. \copyright ESA} 
\label{pancino_fig_rp}
\end{figure}

The resolution is a function of wavelength, ranging from 4 to 32 nm/pix for BP
and 7 to 15 nm/pix for RP. The spectral resolution, R=$\lambda/\delta \lambda$
ranges from 20 to 100 approximately. The dispersers have been designed in such a
way that BP and RP spectra are of similar sizes (45 pixels). Window extensions
meant to measure the sky background are implemented. To compress the amount of
data transmitted to the ground, all the BP and RP spectra -- except for the
brightest stars -- are binned on chip in the across-scan direction, and are
transmitted to the ground as one-dimensional spectra. Figure~\ref{pancino_fig_rp}
shows a simulated RP spectrum, unbinned, before windowing, compression, and
telemetry.

The final data products will be the end-of-mission (or intermediate releases) of
global, combined BP and RP spectra and integrated magnitudes M$_{BP}$ and
M$_{RP}$. Epoch spectra will be released only for specific classes of objects,
such as variable stars and quasars, for example. The internal flux calibration of
integrated magnitudes, including the M$_G$ magnitudes as well, is expected at a
precision of 0.003~mag for G=13 stars, and for G=20 stars goes down to 0.07~mag in
M$_G$, 0.3~mag in M$_{BP}$ and M$_{RP}$. The external calibration should be
performed with a precision of the order of a few percent (with respect to Vega).

\begin{figure}
\resizebox{\hsize}{!}{\includegraphics{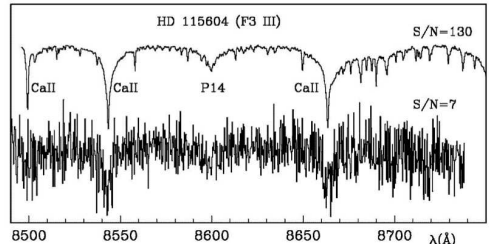}}
\caption{Simulated RVS end-of-mission spectra for the extreme cases of 1 single 
transit (bottom spectrum) and of 350 transits (top spectrum). \copyright ESA}
\label{pancino_fig_rvs}
\end{figure}

\subsection{High-resolution spectroscopy}
\label{sec-rvs}

The primary objective of the RVS is the acquisition of radial velocities, which
combined with positions, proper motions, and parallaxes will provide the means to
decipher the kinematical state and dynamical history of our Galaxy.

The RVS will provide the radial velocities of about 100--150 million stars up to
17-th magnitude with precisions ranging  from 15 km s$^{-1}$ at the faint end, to
1 km s $^{-1}$ or better at the bright end. The spectral resolution,
R=$\lambda$/$\delta\lambda$ will be 11\,500. Radial velocities will be obtained
by cross-correlating observed spectra with either a template or a mask. An
initial estimate of the source atmospheric parameters will be used to  select 
the  most appropriate template or mask. On average, 40 transits will be collected
for each object during the 5-year lifetime of Gaia, since the RVS does not cover
the whole width of the Gaia AF (Figure~\ref{pancino_fig_foc}). In total, we
expect to obtain some 5 billion spectra (single transit) for the brightest stars.
The analysis of this huge dataset will be complicated, not only because of the
sheer data volume, but also because the spectroscopic data analysis relies on the
multi-epoch astrometric and photometric data. 

The covered wavelength range (847-874 nm, Figure~\ref{pancino_fig_rvs}) is a rich
domain, centered on the infrared calcium triplet: it  will not only provide
radial velocities, but also many stellar and interstellar diagnostics. It has
been selected to coincide with the energy distribution peaks of G and K type
stars, which are the most abundant targets. In early type stars, RVS spectra may
contain also weak Helium lines and N, although they will be dominated by the
Paschen lines. The RVS data will effectively complement the astrometric and
photometric observations, improving object classification. For stellar objects,
it will provide atmospheric parameters such as effective temperature, surface
gravity, and individual abundances of key elements such as Fe, Ca, Mg, Si for
millions of stars down to G$\simeq$12. Also, Diffuse Intertellar Bands (DIB)
around 862 nm will enable the derivation of a 3D map of interstellar reddening. 

\section{The DPAC}
\label{sec-dpac}

ESA will take care of the satellite design, build and testing phases, of launch
and operation, and of the data telemetry to the ground, managing the ESAC
datacenter in Madrid, Spain. The data treatment and analysis is instead
responsibility of the European scientific community. In 2006, the announcement of
opportunity opened by ESA was successfully answered by the Data Processing and
Analysis Consortium (DPAC), a consortium that is presently counting more than 400
scientists in Europe (and outside) and more than 25 scientific institutions. 

The DPAC governing body, or executive (DPACE) oversees the  DPAC activities and 
the work has been organized among a few Coordination Units (CU) in charge of
different aspects of data treatment:

\begin{itemize}
\item{{\bf CU1. System Architecture} (manager:  O' Mullane), dealing with all
aspects of hardware and software, and coordinating  the framework for software
development and data management.}
\item{{\bf CU2. Data Simulations} (manager: Luri), in charge of the simulators of
various stages of data products, necessary for software development and testing.}
\item{{\bf CU3. Core processing} (manager: Bastian), developing the main pipelines
such as IDT, AGIS and astrometry processing in general.}
\item{{\bf CU4. Object Processing} (managers: Pourbaix/Tanga), for the processing
of objects that require special treatment such as minor bodies of the Solar
system, for example.}
\item{{\bf CU5. Photometric processing} (manager: van Leeuwen), dedicated to the
BP, RP, and M$_G$ processing and calibration, including image reconstruction,
background treatment, and crowding treatment, among others.}
\item{{\bf CU6. Spectroscopic Processing} (managers: Katz/Cropper), dedicated to
RVS processing and radial velocity determination.}
\item{{\bf CU7. Variability Processing} (managers: Eyer/Evans/Dubath), dedicated
to processing, classification and parametrization of variable objects.}
\item{{\bf CU8. Astrophysical Parameters} (managers: Bailer-Jones/Thevenin),
developing object classification software and, for each object class, software for
the determination of astrophysical parameters.}
\item{{\bf CU9. Catalogue Production and Access} (to be activated in the near
future), responsible for the production of astrophysical catalogues and for the 
publication of Gaia data to the scientific community. }
\end{itemize}

These are flanked by a few working groups (WG) that deal with aspects that are
either transversal among the various CUs (such as the GBOG, coordinating the
ground based observations for the external calibration of Gaia) or  of general
interest (such as the Radiation task force, serving as the interface between DPAC
and the industry in all matters related to CCD radiation tests). 

\subsection{The Italian contribution}

Italian efforts for the preparation of Gaia -- and the participation to the DPAC
--  are regulated by an INAF-ASI agreement with ESA. A coordination group (P.I.
M.~Lattanzi) manages the funding and activities, and contains representatives of
the main Italian institutes involved in DPAC activities. In total, approximately
60 Italian scientists work in the DPAC for a total of 36 FTE, who are involved in
almost all of the CUs activities outlined in the previous Section. 

To summarize the main contributions, we firstly mention the most numerous group,
involved in CU3 and also CU2 activities (INAF-OATO, University of Torino,
Politecnico of Torino, University of Padova), dealing mainly with astrometry and
fundamental physics. There is also a large group involved in CU5 (INAF-OABO,
University of Bologna, INAF-ROMA, INAF-OATE, University of Roma, INAF-OAPD) that
is responsible for all aspects of the external flux calibration of Gaia
photometry and for other aspects of photometric processing such as crowding
issues and image reconstruction. Finally Italian scientists are also involved in
CU7 (INAF-OACT, INAF-OABO, INAF-OANA) working on algorithms for the
classification and parametrization of particular kinds of variable objects
(mainly stellar pulsators), and in CU8 (INAF-OAPD, University of Padova,
INAF-OANA, INAF-OACT) dealing mainly with spectral libraries for the
classification and parametrization of stellar sources. Apart from these four main
groups, there are also smaller groups or individuals that are DPAC active members
on other topics.

All in all, Italy is one of the major DPAC contributors, both in terms of funding
and in terms of FTE and responsibilities.

\bibliographystyle{aa}

\begin{thebibliography}{}

\bibitem[Ibata \& Gibson(2007)]{ibata07} Ibata, R., \& Gibson, B.\
2007, Scientific American, 296, 040000 

\bibitem[Perryman et al.(1997)]{perryman97} Perryman, M.~A.~C.,  Lindegren, L.,
\& Turon, C.\ 1997, Hipparcos - Venice '97, 402, 743 



\end{thebibliography}

\end{document}